\documentclass[10pt]{article}

\usepackage[utf8]{inputenc} 

\usepackage{geometry} 
\usepackage{graphicx} 
\usepackage[parfill]{parskip} 
\usepackage{booktabs} 
\usepackage{array} 
\usepackage{paralist} 
\usepackage{verbatim} 
\usepackage{subfig} 
\usepackage{authblk}
\usepackage{amssymb,amsmath}
\usepackage{color}

\usepackage{fancyhdr} 
\usepackage{sectsty}
\allsectionsfont{\sffamily\mdseries\upshape} 

\def\kv{\kappa_V}
\def\kl{\kappa_\ell}
\def\ku{\kappa_u}
\def\kd{\kappa_d}

\def\ifb{fb$^{-1}$}
\def\iab{ab$^{-1}$}

\newcommand{\be}{\begin{equation}}
\newcommand{\ee}{\end{equation}}
\newcommand{\bea}{\begin{eqnarray}}
\newcommand{\eea}{\end{eqnarray}}
\newcommand{\bi}{\begin{itemize}}
\newcommand{\ei}{\end{itemize}}

\title{Discriminators of 2 Higgs Doublets at the \\LHC14, ILC and MuonCollider(125): \\A Snowmasss White Paper}
\author[1]{Vernon~Barger}
\author[1]{Lisa~L.~Everett}
\author[2]{Heather~E.~Logan}
\author[1]{Gabe~Shaughnessy}
\affil[1]{Department of Physics, University of Wisconsin, Madison, WI 53706, USA}
\affil[2]{Ottawa-Carleton Institute for Physics, Carleton University, Ottawa, ON K1S 5B6, Canada}

\begin{document}

\maketitle
\begin{abstract}
The historic LHC discovery of the 125 GeV particle with properties that closely resemble the Standard Model (SM) Higgs boson verifies our understanding of electroweak symmetry breaking, but solidifies the need for a resolution to the hierarchy problem.  Many extensions of the SM that address the hierarchy problem contain a non-minimal Higgs sector.  Therefore, as a benchmark alternative to the SM Higgs mechanism, we study a general 2 Higgs doublet model (2HDM-G) framework for evaluating future sensitivity to Higgs couplings.  We study how well it can be distinguished from the SM Higgs boson by future measurements at LHC14, ILC (250, 500,1000 GeV) and a Muon Collider (125 GeV).  Additionally, our study bears on singlet Higgs extensions of two Higgs doublet models through predicted coupling relationships.

\end{abstract}


The similarity to the SM Higgs boson of the 125 GeV particle discovered at LHC7 and LHC8  makes future precision measurements of its couplings  of paramount importance.  The SM Higgs mass receives large radiative corrections that grow quadratically with an ultra-violet cutoff.  A possible resolution to the ensuing problem from the large hierarchy between the electroweak and Planck scales is new physics at the TeV scale, such as supersymmetry, extra-dimensions or new top-like states, to cancel the SM contributions. Further, measurements of its gauge couplings can  determine whether it is solely responsible for the spontaneous breaking of the electroweak $SU(2)\times U(1)$ symmetry.  

Any new physics scenario should lead to changes in the SM predictions of the lightest neutral CP-even Higgs particle.  In ascertaining the extent that future collider measurements may detect deviations from SM couplings, an alternative EWSB mechanism can be chosen as a benchmark.  In our study we adopt a general 2 Higgs doublet model (2HDM-G), in which both Higgs doublets can couple to to the down-type quarks and to the charged-leptons, to encapsulate observable deviations from a SM Higgs.  This is a commonly chosen new physics model, even though it does not provide a solution the hierarchy problem per se.


Our considerations include (i) the energy upgrade of the LHC to 14 TeV with an integrated luminosity  300 \ifb and 3 ab$^{-1}$~\cite{Peskin:2012we},  (ii) an International Linear Collider (ILC) that would successively operate at 250, 500 and 1000 GeV and accumulate integrated luminosities of 200, 500 and 1000 fb$^{-1}$ respectively~\cite{Baer:2013cma}, or  (iii) a Muon Collider (MC) that would scan over the $s$-channel Higgs resonance~\cite{Barger:1995hr}; the Compact Linear Linear Collider (CLIC) option at 500 GeV and higher energy would give similar physics results as the ILC for comparable integrated luminosities.  We perform a fit which uses $\sigma \cdot{\rm BR}$ precisions anticipated at each of the machines along with the recoil mass cross section measured at the ILC.

The upgraded LHC will yield much higher Higgs signals but will also have higher backgrounds, especially from overlapping events.  An entry-level ILC at 250 GeV is specifically  positioned to produce Higgs bosons copiously via $Z$ Higgstrahlung and thereby measure this cross-section to $\approx 2.5\%$~\cite{Baer:2013cma}.  With higher ILC energies,  above $\sqrt s \gtrsim 400$ GeV~\cite{Baer:2013cma}, the Vector Boson Fusion (VBF) cross section becomes the dominant Higgs production channel.  Hence, at $\sqrt s = $ 500 GeV (and at 1 TeV), an independent precision measurement can be made of the vector boson coupling to the Higgs particle.

The general 2HDM includes as special cases the traditional types of 2HDMs: Type-I, Type-II, Lepton-specific, and Flipped~\cite{Barger:2009me}.  In addition to the inclusion of the mixing of the two neutral CP-even Higgs states which impacts the weak boson couplings, the flavor structure of the Yukawa couplings is allowed to be generic.  The Yukawa Lagrangian we that we adopt is~\cite{Barger:2013xx} which is the same physics as the aligned model~\cite{Pich:2009sp}
\be
- {\cal L} = y_u\,\overline{u}_R \, {\bf \Phi}_u \,Q_L    +\,y_d\, \overline{d}_R\,(c_{\gamma_d}\,{\bf \Phi}_d \,+\, s_{\gamma_d}\,\tilde{{\bf \Phi}}_u) \, Q_L  +  y_\ell\,\overline{e}_R \, (c_{\gamma_\ell}\,{\bf \Phi}_d \,+\, s_{\gamma_\ell}\,\tilde{{\bf \Phi}}_u) \,L_L     \,+\, \mbox{h.c.}\,,
\ee
where $\gamma_{\ell}$ and $\gamma_d$ parameterize the two Higgs doublet's coupling to charged-leptons and down-type quarks, respectively.  Measuring the relative values of the Yukawa couplings potentially offers a direct window to the flavor structure of any new physics.  The four independent couplings can be written in terms of the usual 2HDM mixing angles and $\alpha$ and $\beta$ as follows
\be
\kv=s_{\beta-\alpha},\quad \, \ku=c_\alpha /s_\beta, \quad \,
\kd=-s_{\alpha-\gamma_d}/c_{\beta-\gamma_d},\quad \,\kl=-s_{\alpha-\gamma_\ell}/c_{\beta-\gamma_\ell}.
\ee
We calculate the $hgg$ and $h\gamma\gamma$ couplings through loops of SM states, assuming that the contributions of a heavy charged Higgs to the $h\to \gamma\gamma$ loop can be neglected.

We provide the following comparative projections based on our fits described fully in~\cite {Barger:2013xx} for distinguishing 2HDMs from the SM with future data from the LHC, ILC or MC:

\begin{figure}[htbp]
\begin{center}
     \includegraphics[angle=0,width=0.47\textwidth]{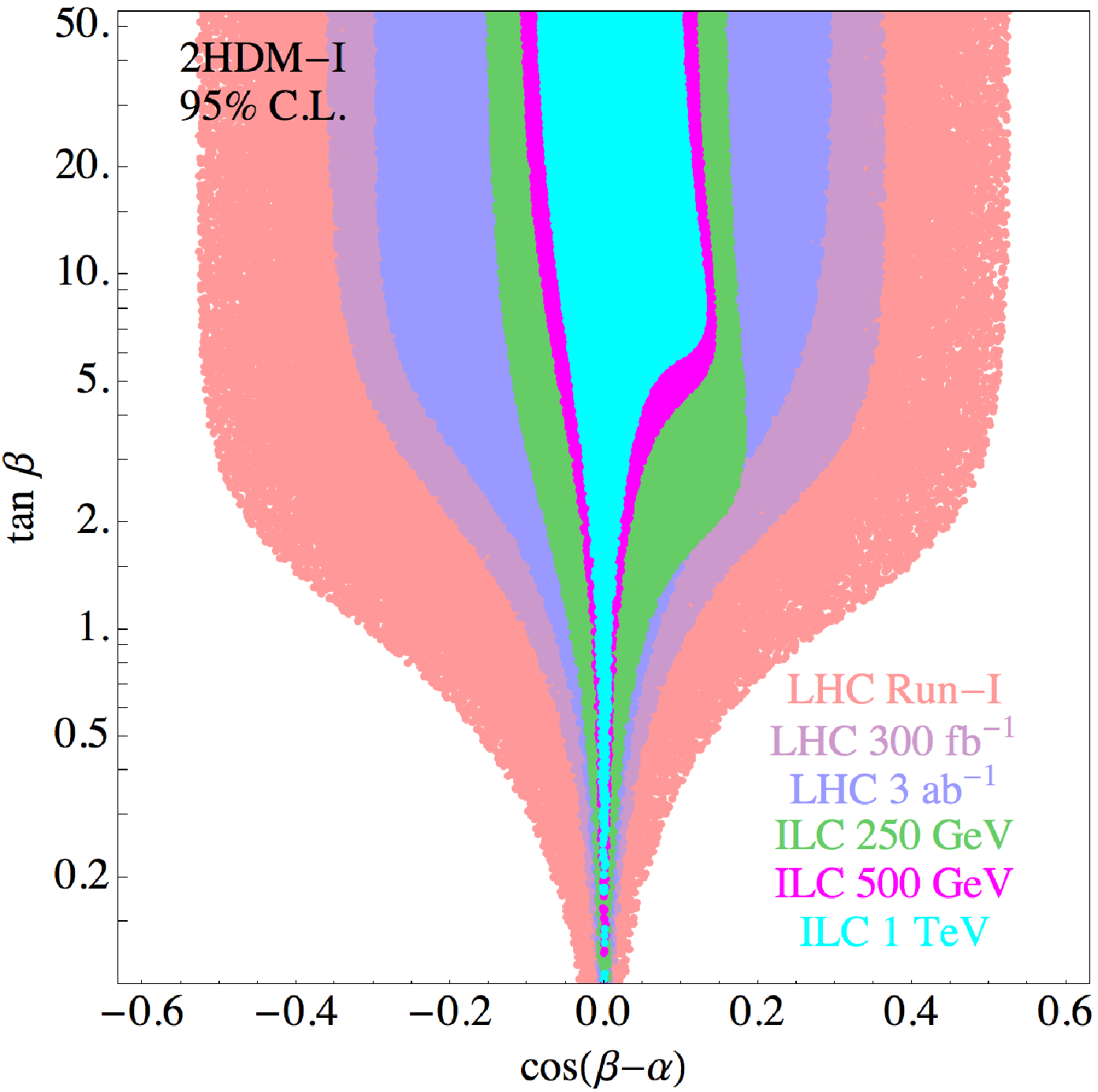}
     \includegraphics[angle=0,width=0.47\textwidth]{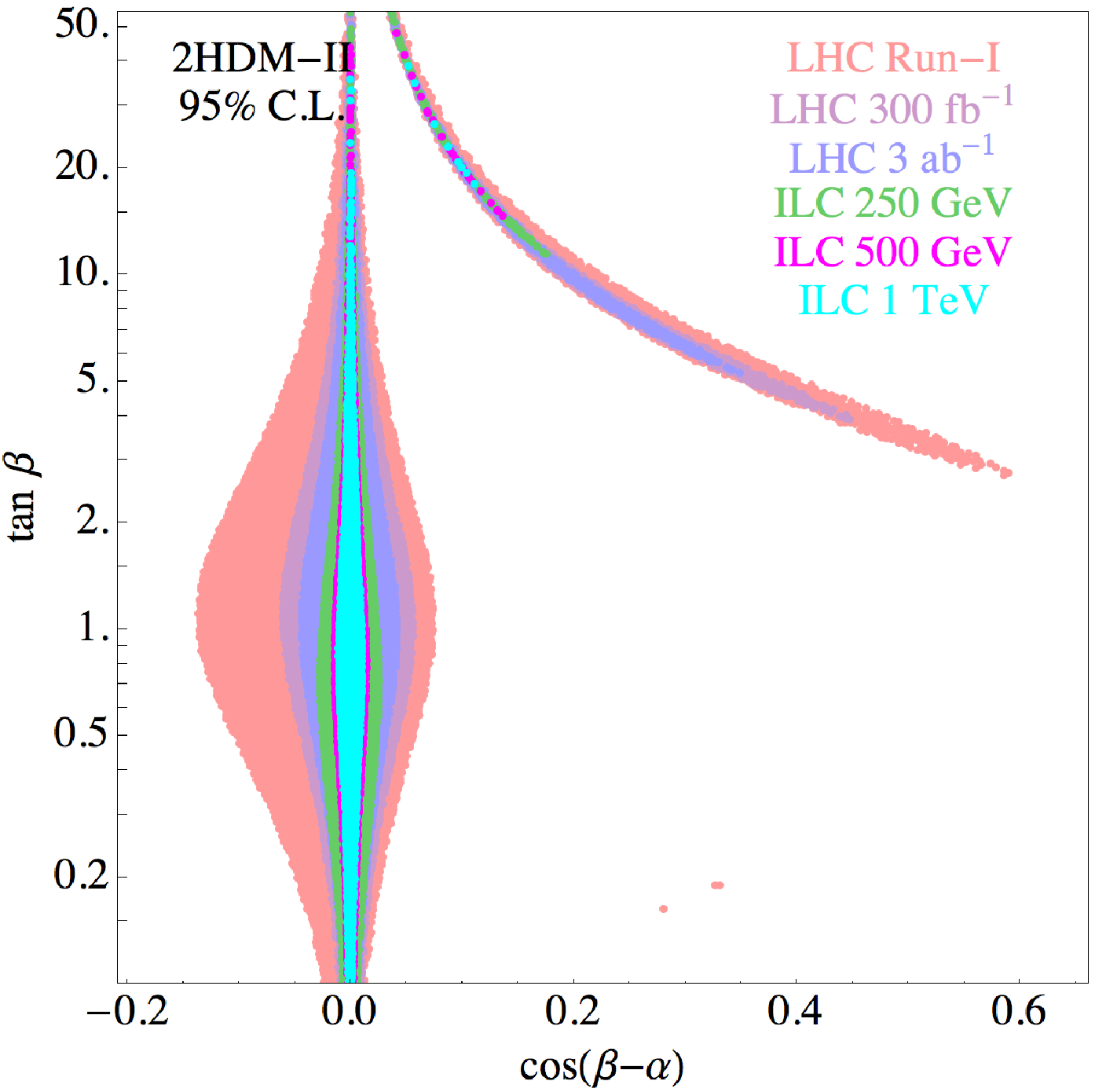}\\
     \includegraphics[angle=0,width=0.47\textwidth]{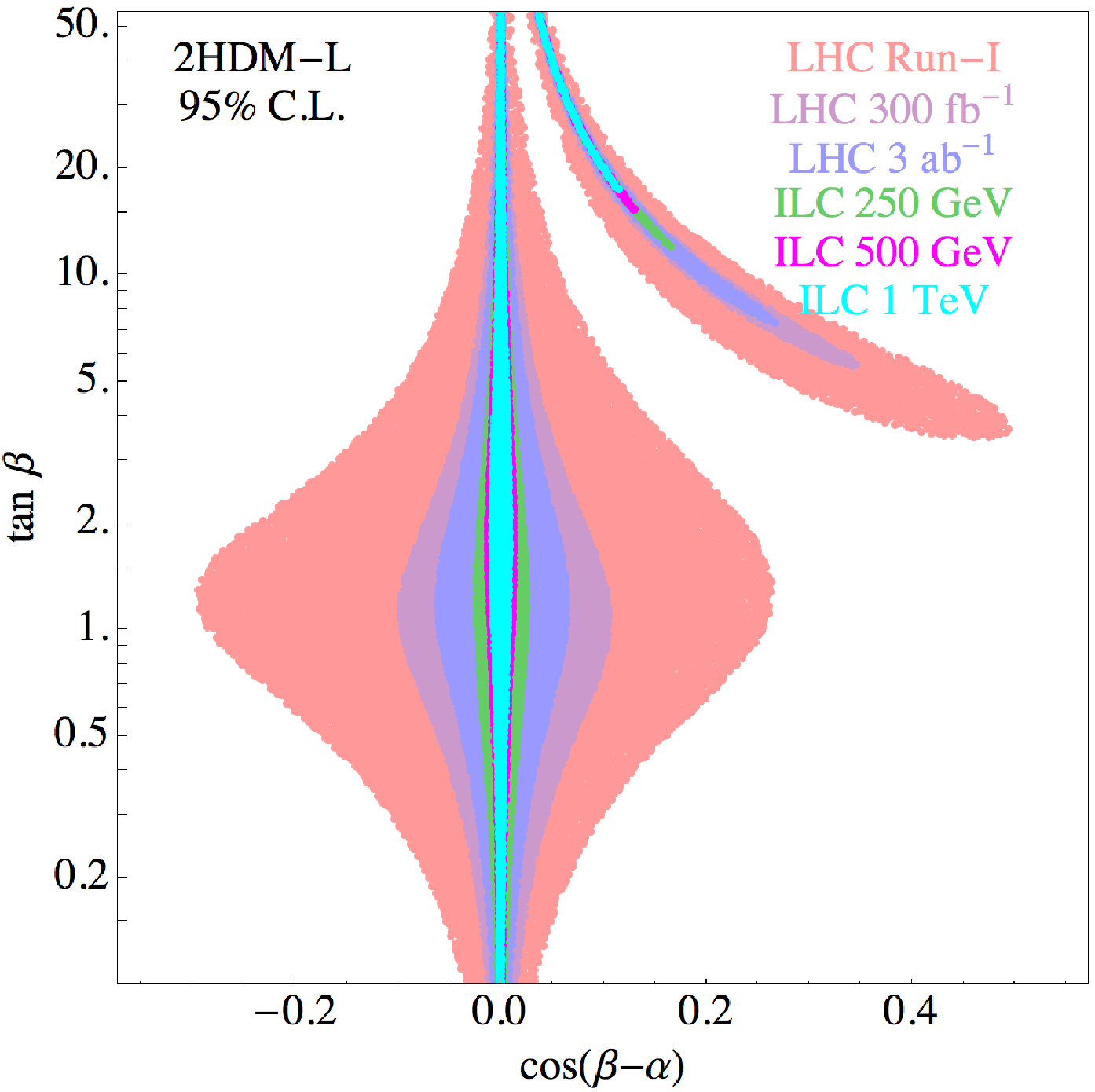}
     \includegraphics[angle=0,width=0.47\textwidth]{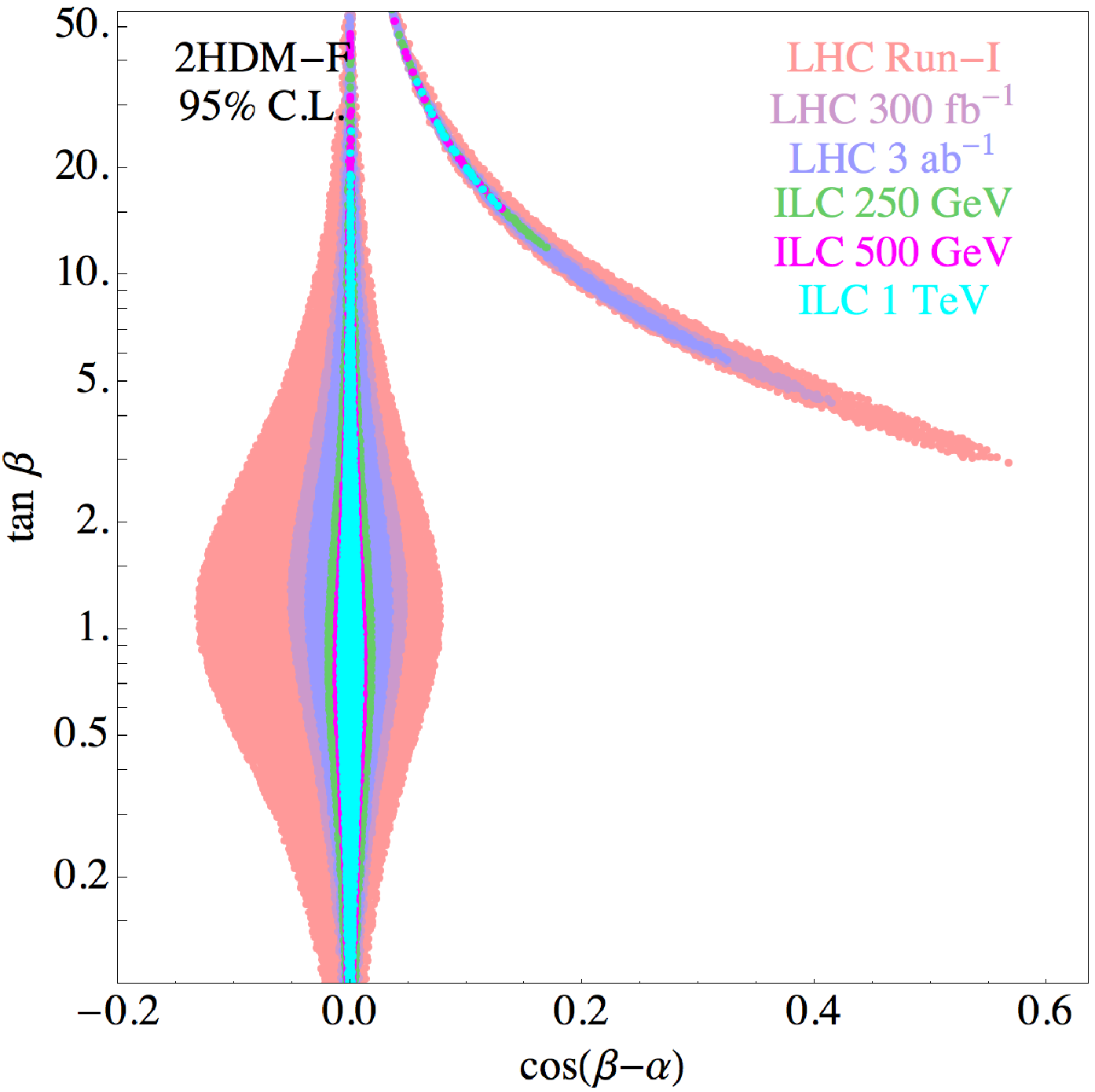}
\caption{The parameter regions at the 95\% C.L. for the usual two Higgs doublet models in the $\tan\beta$ vs. $\cos(\beta-\alpha)$ plane: a) 2HDM-I, b) 2HDM-II, c) 2HDM-L, d) 2HDM-F.  Similar contours may be found in Ref.~\cite{Chen:2013rba}.}
\label{fig:cba-tb}
\end{center}
\end{figure}
\bi
\item The decoupling parameter $\cos(\beta-\alpha)$ can be measured in traditional 2HDM channels to remarkable precision.  A bulk of the $1\sigma$ region lies along $\cos(\beta - \alpha)\sim0$, but a secondary branch extends in the $\tan\beta-\cos(\beta - \alpha)$ plane that flips the sign of $\kd$ and/or $\kl$.  Fig.~\ref{fig:cba-tb} illustrates the regions compatible with a SM-like Higgs boson at the 95\% C.L.

\item The measurement of Yukawa coupling ratios tests the Higgs flavor structure.  Specifically, the LHC14 at 300 \ifb (3 \iab) can measure the $\kd/\ku$ and $\kl/\ku$ coupling ratios to 15\% and 10\%, respectively.  The ILC  250, 500, 1000 can measure these ratios to 4\%, 2\% and 1.5\%, respectively.  Fig.~\ref{fig} illustrates the exclusion level achievable.  

\item The ``wrong Higgs'' couplings that can be induced by new physics loops in 2HDMs can be easily probed at the ILC.   Examples are the gluino and squark loops of SUSY models, which are Type-II. The loop corrections are manifest as shifts in the $b$ and $\tau$ Yukawa couplings.  These shifts deviate away from the lines for the respective models, as indicated in Fig.~\ref{fig}.

\item The total Higgs width can be inferred at the LHC to $\sim 20\%$ if we demand that $\kv < 1$ to be consistent with unitarity.  The ILC250 can measure the total $Zh$ production cross section, thereby fixing the normalization of $\kv$ and thus constrain the total Higgs width.  In our fits, this translates to being sensitive to new physics contributions to Higgs decay of  $\Gamma_X \gtrsim 0.26 $ MeV ($\sim6\%$ branching fraction) at the 95\% C.L..

\item The existence of Higgs singlets in addition to the two Higgs couplets would reduce the gauge and Yukawa couplings, while extra doublets can reduce or enhance Yukawa couplings.  Pattern relations among the gauge and Yukawa couplings can help establish the underlying model~\cite{Barger:2009me}.  At the ILC, the singlet-Higgs mixing angle can be probed to $\sin\theta \lesssim 0.1$ at the 95\% C.L.  Therefore, the LHC14 upgrade, the ILC and the MC are all well positioned to probe and distinguish these mixing scenarios from a SM Higgs.

\begin{figure}[htbp]
\begin{center}
     \includegraphics[angle=0,width=0.47\textwidth]{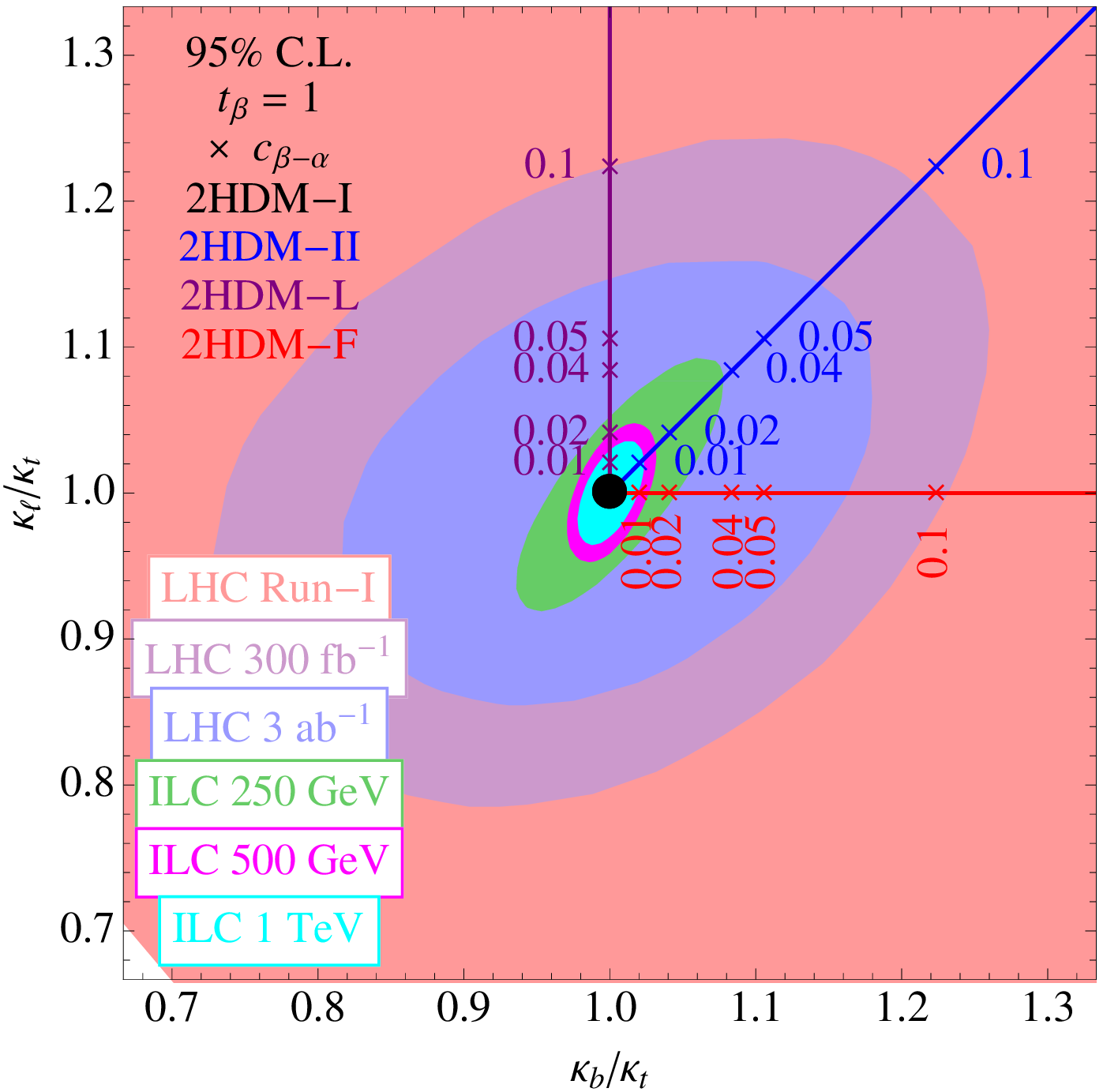}
     \includegraphics[angle=0,width=0.47\textwidth]{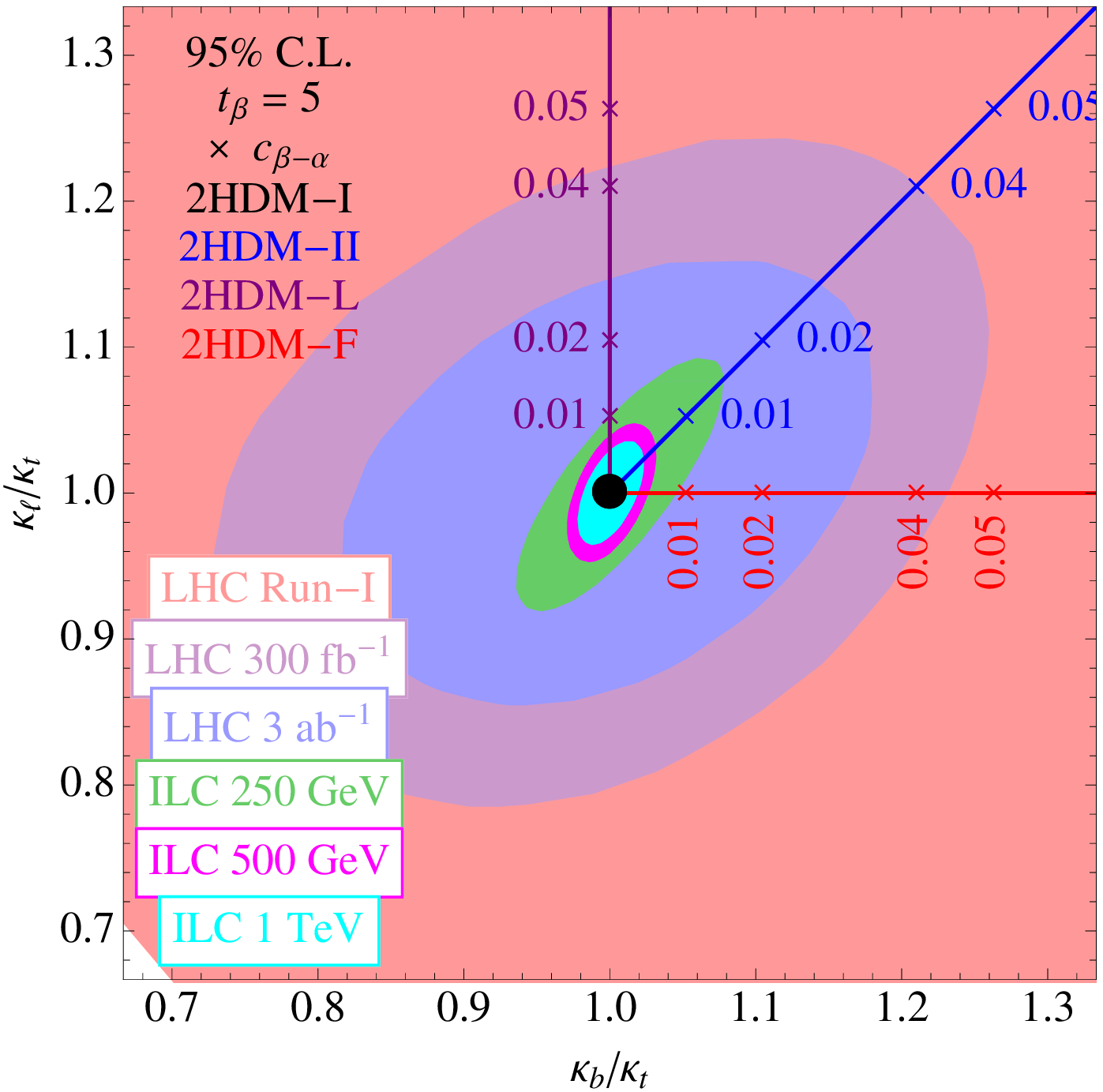}\\
     \includegraphics[angle=0,width=0.47\textwidth]{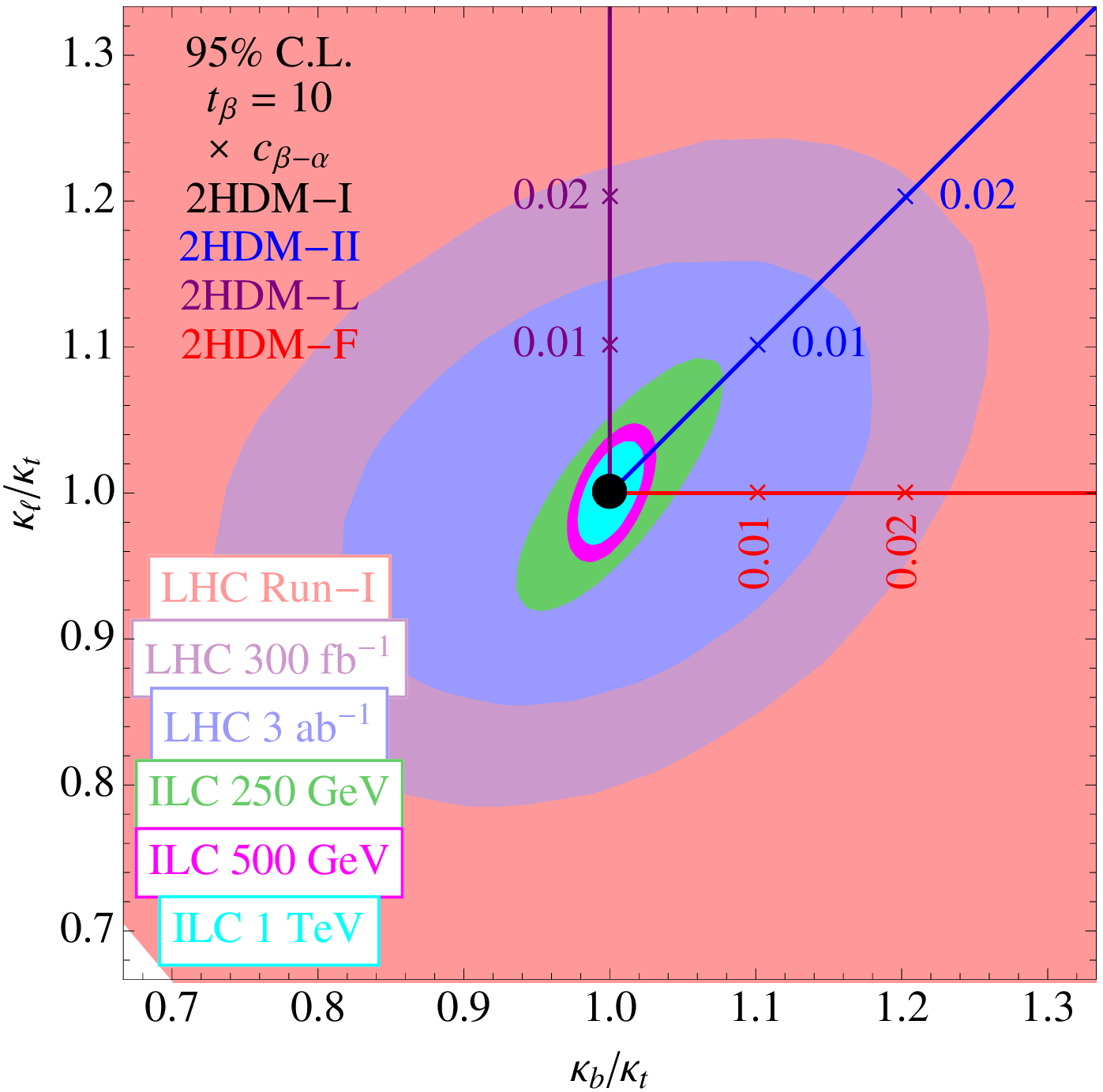}
     \includegraphics[angle=0,width=0.47\textwidth]{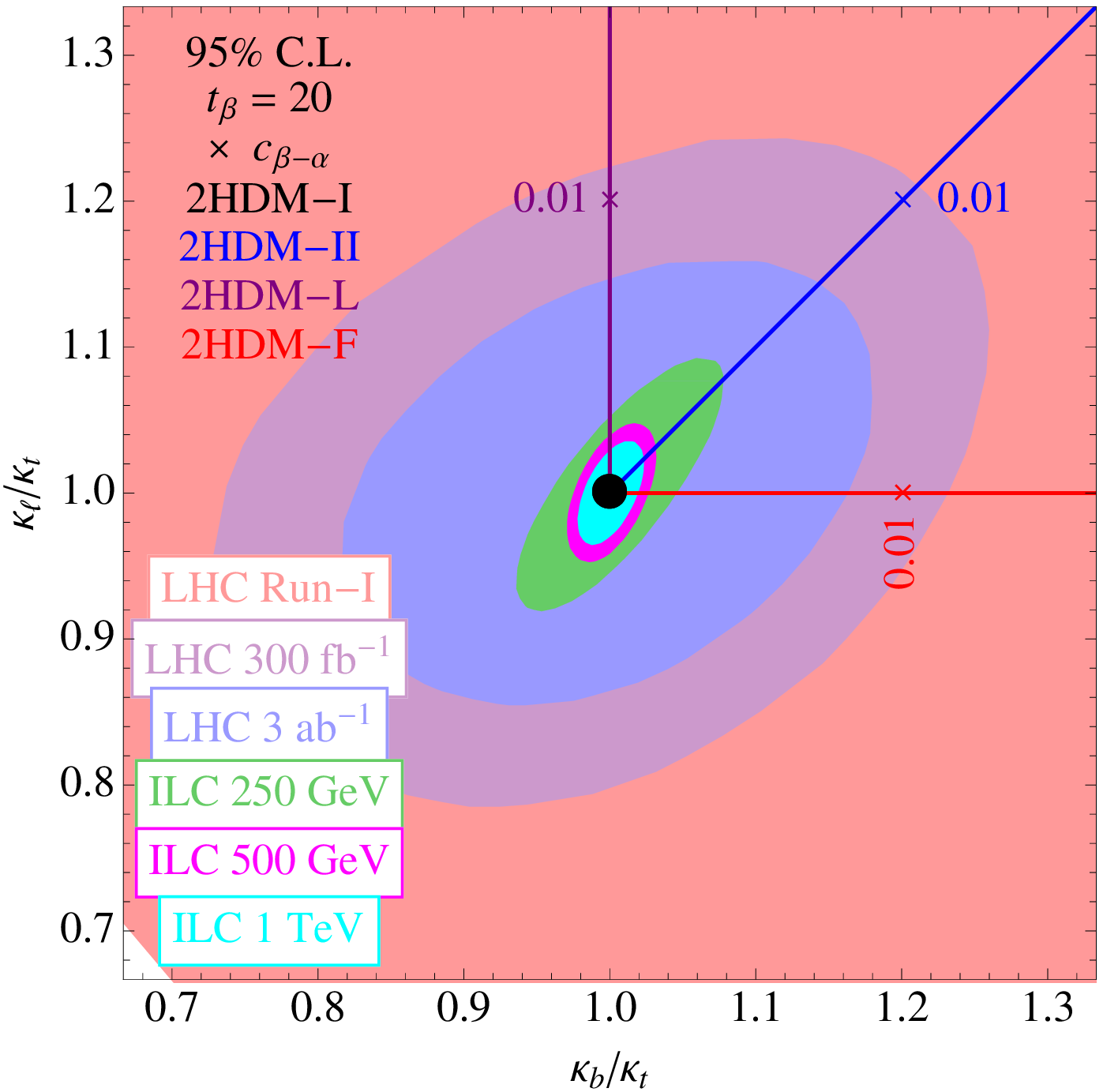}
\caption{Yukawa coupling ratios at the 95\% C.L.  from current data and simulated LHC14 and ILC data with the expected 2HDM predictions, at given representative values of $\tan \beta$. The black dot marks the 2HDM-I model, which in this plane matches the SM.  Figures taken from Ref.~\cite{Barger:2013xx}}
\label{fig}
\end{center}
\end{figure}

\ei

In addition, we note the following characteristics of the collider programs:
\bi
\item The ILC can probe invisible decays of the Higgs boson via the recoil spectrum of the associated $Z$ boson in $e^+e^-\to Zh$ production.  The anticipated measurement uncertainty of the Higgs invisible decay is ${\cal O}(0.5\%)$~\cite{Peskin:2012we}.  This corresponds to an upper limit on $\Gamma_{\rm inv}$ of 0.04 MeV at 95\% C.L.
Any decays of the Higgs boson to new physics channels can also be constrained by the unit sum of the branching fractions obtained in the recoil tagged events.  At ILC1000, the uncertainty on the total width is 2.4\% at the 95\% C.L.~\cite{Baer:2013cma}.

\item A Muon Collider tuned to the Higgs mass can scan over the Higgs resonance profile.  This is the only way to directly measure the total Higgs width.  This is anticipated to yield an  uncertainty down to the percent level in the total width~\cite{Han:2012rb,Cline:2013xx}.    In turn, this determination provides sensitivity to new physics decays contributions of $\Gamma_X \gtrsim 0.05 $ MeV at the 95\% C.L., a factor of 2 better than the ILC1000 determination based on $\sigma_{Zh}$ alone.  The MC can determine the Higgs coupling to muons to an accuracy of 1 percent relative to the SM.

\ei

\section{Acknowledgements}
VB and LE thank KITP-UCSB for kind hospitality during the progress of this work. This research was supported in part by the National Science Foundation under Grant No. NSF PHY11-25915. VB, LE, and GS are supported by the U. S. Department of Energy under contract DE-FG-02-95ER40896.  H.E.L. was supported by the Natural Sciences and Engineering Research Council of Canada.


\end{document}